\documentclass[aps,prl,showpacs,amssymb,superscriptaddress, notitlepage,floats,floatfix,nofootinbib, twocolumn]{revtex4-1}

\usepackage{graphicx}
\usepackage{amssymb}
\usepackage{amsmath}
\usepackage{amsfonts}
\usepackage{bm}
\usepackage{mathrsfs}
\usepackage[colorlinks=true]{hyperref}
\usepackage[all]{hypcap}
\usepackage[utf8]{inputenc}
\usepackage{color}
\usepackage{multirow}
\usepackage{palatino}
\usepackage{float}
\usepackage{soul}
\usepackage{calrsfs}

\begin{document}
\title{Spectral Lines of Quantized, Spinning Black Holes and their Astrophysical Relevance}
\author{Andrew Coates}
\email{andrew.coates@uni-tuebingen.de}
\author{Sebastian H. V\"olkel}
\email{sebastian.voelkel@uni-tuebingen.de}
\author{Kostas D. Kokkotas}
\affiliation{Theoretical Astrophysics, IAAT, University of T\"{u}bingen, 72076 T\"{u}bingen, Germany}
\date{\today}
%
\begin{abstract}
In this Letter, we study black hole area quantization in the context of gravitational wave physics. It was recently argued that black hole area quantization could be a mechanism to produce so-called echoes as well as characteristic absorption lines in gravitational wave observations of merging black holes. One can match the spontaneous decay of these quantum black holes to Hawking radiation calculations. Using some assumptions, one can then estimate the natural widths of these states. As can be seen from a classical paper by Bekenstein and Mukhanov, the ratio between width and spacing of nonspinning black hole states approaches a small constant, which seems to confirm the claim. However, we find that, including the effect of black hole spin, the natural widths increase. To properly address any claim about astrophysical black holes, one should examine the spinning case, as real black holes spin. Thus, the word “spinning” is key to the question of whether or not black holes should have an observable spectrum in nature. Our results suggest that it should be possible to distinguish between any scenarios for which the answer to this question is yes. However, for all of the commonly discussed scenarios, our answer is “almost certainly no.”

%

%
\end{abstract}
\maketitle
\section{Introduction}\label{Introduction}
Introduction.—The quantum nature of black holes is among the most fascinating puzzles in theoretical physics. A particularly interesting hint about the nature of black holes comes from the seminal works on the thermodynamics of black holes \cite{PhysRevLett.26.1344,Bekenstein1972,PhysRevD.7.2333,Bardeen1973,Hawking:1974rv,Hawking1975,PhysRevLett.57.397}. As we have no generally accepted theory of quantum gravity, nor any experimental access, following up these calculations is very challenging. 
\par
On the other hand there is a growing interest in exotic compact objects (ECOs), which could potentially mimic black holes in most astrophysical situations, but have some fundamental differences. Many different models have been proposed, among them are ultracompact constant density stars \cite{Chandrasekhar:1991fi}, gravastars \cite{2001gr.qc.....9035M,Visser:2003ge}, and the many faces of wormholes \cite{Bronnikov:1973fh,Ellis:1973yv}. There are also proposed modifications of the physics on the horizon scale of classical black holes \cite{2017JHEP...05..054B}, e.g. firewalls \cite{Almheiri:2012rt}. 
\par
One way these ECOs often differ from black holes is the appearance of so-called gravitational wave echoes \cite{Cardoso:2016rao,Cardoso:2016oxy,Cardoso:2017cqb}, which would appear after binary mergers and other astrophysical events in some of the scenarios. This effect on the time evolution was first reported for ultracompact, constant density stars where some or all of the incoming gravitational radiation would be reflected in such systems \cite{Kokkotas:1995av,PhysRevD.60.024004,2000PhRvD..62j7504F}. This radiation then bounces between the would-be horizon and the potential barrier for perturbations and excites trapped quasinormal modes. For every bounce, some part of the radiation can escape, producing repeated bursts of radiation, as seen by an external observer. The claim that tentative evidence was found in LIGO detections \cite{2017PhRvD..96h2004A,2017arXiv170103485A,Conklin:2017lwb} triggered a lot of work in the field, see \cite{Cardoso:2017cqb} and references therein. However, these claims are highly disputed \cite{2016arXiv161205625A,2017arXiv171209966W} and more precise future gravitational wave detections are required to clarify the situation. As of April 2019, the LIGO and Virgo observatories have entered the O3 observing run with increased sensitivity.
\par
Our Letter was inspired by the recent claims \cite{Foit:2016uxn,Cardoso:2019apo} that black hole area quantization, as proposed by Bekenstein and Mukhanov \cite{Bekenstein:1974jk,Mukhanov:1986me,Bekenstein:1995ju}, leads to gravitational wave echoes. (As the name suggests, in the area quantization scenario, the area of black holes is quantized in some unit area of the same order as the Planck area) However, one might naively expect that, for astrophysical black holes, any change due to area quantization should be washed out. This is particularly the case if one expects to recover general relativity in the classical limit. Thus the prediction of echoes should only be relevant for small black holes, not for astrophysical ones. In order to quantify this line of thought, we extend the preexisting estimates of the natural linewidths \cite{Bekenstein:1995ju} to the spinning case. The method matches the spontaneous decay of quantized black holes to the power output of Hawking radiation. For large, nonspinning black holes, there is an almost universal ratio of the width to the gap between neighboring states. This width is small in all proposed scenarios. This means that well-isolated, nonspinning black holes are likely good reflectors, if their areas are quantized. For spinning black holes, the situation is a little less clear. First of all there are more decay channels and so one would expect shorter lifetimes and hence large widths. (Specifically, the lifetime of the individual states should be shorter, not the total lifetime of the black holes.) Second, for large enough spins, the area actually grows via Hawking radiation \cite{Page:1976ki}. This is a consequence of superradiance and the fact that the surface gravity is small for large \(a\) (see, e.g. \cite{Wald:1984rg}). These also provide the counter-intuitive behaviour that the power output grows with spin, even as the temperature decreases. We will see that the conclusion in the spinning case depends on the scale of the area quantization. In particular, for the more common suggestions this leads to no observable effect, but if the quantum of area is on the large side there will be. They should also be distinguishable where there is a detectable effect.
\par
Recently, other works appeared \cite{Oshita:2019sat, Wang:2019rcf}, tackling this problem from a different direction (see the discussion).
\section{Widths of non-spinning states}\label{widths}
For demonstrative purposes, we review the classic calculation by Bekenstein and Mukhanov in \cite{Bekenstein:1995ju} the premise of which is that we can make an estimate of the natural width of nonspinning states using various assumptions:
\begin{enumerate}\setcounter{enumi}{-1}
\item Isolated black holes can only have areas an integer multiple of some unit of area.
\item The correct definition of the energy of the black hole is just \(M c^2\).
\item For large black holes, the semiclassical Hawking radiation should give the correct power emitted in spontaneous decays.
\item Spontaneous decays that change the spin of the black hole by a large amount are highly suppressed. In other words, the decay of a Schwarzschild black hole should not give it a significant spin.
\end{enumerate}
 Assumption ($0$), that the area of a Schwarzschild black hole is quantized, can be written as
\begin{align}\label{eq-area}
A
= \alpha \ell_p^2 n 
= 4 \pi r_\text{s}^2 
= 4 \pi \left(\frac{2G M}{c^2}\right)^2.
\end{align}
Here we have the Planck length $\ell_p^2 = \hbar G/c^3$ and the Schwarzschild radius $r_\text{s}$. The integer $n$ labels the state of the black hole. The constant $\alpha$ is \textit{a priori} not known and depends on the specific model of area quantization. The different proposals for $\alpha$ made in the literature range from \(4 \ln 2\) to \(32 \pi\) or even larger \cite{Bekenstein:1974jk,Bekenstein:1995ju,Mukhanov:1986me,Hod:1998vk,Dreyer:2002vy,Maggiore:2007nq,Louko:1996md}.  (In principle \(4 \ln q\) for integer \(q>2\) is included in the literature but \(q<10\) has been the focus.)
\par
We now invert Eq. \eqref{eq-area} to find the mass of the \(n\mathrm{th}\) state, $M_n$, which gives us
\begin{align}
M_n 
= \sqrt{\frac{\alpha n c\hbar}{16 \pi G}}.
\end{align}

So, if we say that assumption ($1$), the energy of the $n$ state is $M_n c^2$, then the transition from $n$ to $n-\delta n$ has an energy 
\begin{align}
\left(M_n - M_{n-\delta n}\right)c^2 
= \hbar \omega_{n,\delta n}.
\end{align}
One can justify this by considering the relation between the Bondi mass and outgoing radiation fluxes \cite{Wald:1984rg}.
\par
To be able to make any quantitative predictions, we need to know the widths of these states. To do so would usually require knowledge of how these states interact with external fields etc.; in other words we would need a theory of quantum gravity. We can, however, make an estimate if we take assumption ($2$), i.e., that the power output by the spontaneous decay of these states matches the power output of Hawking radiation. In this case, we would have
\begin{align}
P_\mathrm{H} 
=\frac{\left<E_\mathrm{emitted}\right>}{\tau},
\end{align}
where $P_\mathrm{H}$ is the power emitted in Hawking radiation, $\left<E_\mathrm{emitted}\right>$ is the average energy emitted in a spontaneous decay, and $\tau$ is the lifetime of the state. We then apply the standard relation $\tau = \hbar/\Gamma$ where $\Gamma$, is the energy width of the state. This gives us
\begin{align}
P_\mathrm{H}=\left<\omega\right>\Gamma,
\end{align}
where $\left<\omega\right>$ is the average transition frequency
\begin{align}
\left<\omega\right>\approx \sum_{\delta n=1}^{n}p(n\to n-\delta n) \omega_{n,\delta n},
\end{align}
with $p(n \to m)$ being the probability that the state will decay from the $n$ state to the $m$ state, given that a decay happens. [This uses the fact that, for large \(n\) and small spin parameter \(a\), small changes in angular momentum do not make a substantial change to the transition frequency, that \(\delta n =0\) transitions increase the mass for nonspinning black holes, and  assumption (\( 3\)).] Importantly, this average can never be smaller than $\omega_{n,1}$, which is the minimum possible transmission frequency. Underestimating $\left<\omega\right>$ will overestimate $\Gamma$. Overestimating $P_\mathrm{H}$ also overestimates $\Gamma$. So if we replace  $\left<\omega\right>$ by $\omega_{n,1}$ and neglect gray body factors in $P_\mathrm{H}$ we will only overestimate $\Gamma$. This likely overestimate of the width is
\begin{align}
\Gamma < \tilde{\Gamma} 
=& A_\text{eff} \sigma_\mathrm{SB} T_\mathrm{H}^4\left(\omega_{n,1}\right)^{-1},
\end{align}
where
\begin{align}
A_\text{eff} &= 4\pi \left( \frac{ 3\sqrt{3}  G M(n)}{c^2}\right)^2, \quad
\sigma_\mathrm{SB} = \frac{ \pi^2 k_\mathrm{B}^4}{60 c^2\hbar^3 },\quad \\
T_\mathrm{H} &= \frac{\hbar c^3}{8 \pi G M(n) k_\mathrm{B}},\quad
\omega_{n,1} = \frac{\left(M(n)-M(n-1)\right)c^2}{\hbar}.
\end{align}
\(A_\text{eff}=  27 A/4\) is based on estimates of the power in Hawking radiation \cite{PhysRevD.13.198}, \(\sigma_\mathrm{SB}\) is the Stefan-Boltzmann constant, and \(T_\mathrm{H}\) is the Hawking temperature of the black hole. Tidying up and expanding for large $n$ we arrive at
\begin{align}
\tilde{\Gamma} =& 9 \frac{\sqrt{n}+\sqrt{n-1}}{320 n}\sqrt{\frac{c^5 \hbar \pi }{G \alpha^3}}
\\
=&\frac{9}{160}\sqrt{\frac{\pi \hbar c^5 }{G \alpha^3 n}}\left(1-\frac{1}{4n}\right)+\mathcal{O}(n^{-5/2}).
\end{align}
This shows that this estimate of the width goes to zero for large black holes (large $n$). Although this is in contradiction with the intuition gained from usual quantum systems, one can see that this must be the case. The result is a basic consequence of the fact that the total lifetime $\tau_\mathrm{tot}$ of a black hole, according to Hawking radiation, scales like \(M^3\) and thus, in the case of area quantization, \(n^{3/2}\). However,
\begin{align}
\tau_\mathrm{tot}
=\sum_{n=0}^N \tau_n\propto\sum_{n=0}^N\frac{1}{\Gamma_n},
\end{align}
and so if $\Gamma_n$ were to be a constant in $n$, then the lifetime would be $\tau_\mathrm{tot}\propto N \propto M^2$, so $\Gamma_n$ must go as $1/\sqrt{n}$. We can get a feel for how large this width is by looking at the ratio
\begin{align}\label{eq-ratio}
\frac{\tilde{\Gamma}}{\hbar\omega_{n,1}}
=\frac{9 \pi}{20 \alpha^2}\left(1-\frac{1}{2 n}+\mathcal{O}(n^{-2})\right).
\end{align}
Note that this approaches a constant for large black holes.
The most common proposals for $\alpha$ are $ \alpha = 4 \ln(q)$ \cite{Bekenstein:1974jk,Bekenstein:1995ju,Dreyer:2002vy}, with $q=2,3$, $\alpha = 8\pi$, or even as high as $\alpha = 32 \pi$ \cite{Louko:1996md}. These give ratios of around \(0.2\), \(0.1\), \(10^{-3}\), and \(10^{-4}\), respectively. Without broadening all of these values would thus lead to very different behavior from the classical black hole.
\par
However, as already mentioned, spin can change the picture considerably \cite{Page:1976ki}. For example, by looking at Fig. 3 of the same work one might expect that including the spin will reduce the lifetime of individual states.
\section{Including the Spin}
The preceding section was limited to the idealized case of nonrotating black holes. However, the most important astrophysical black holes, from an observational point of view, are expected to have non-negligible spin. Robust estimates for binary black hole mergers, which match with the LIGO and Virgo detections \cite{PhysRevLett.116.061102,PhysRevLett.116.241103,PhysRevLett.118.221101,PhysRevLett.119.141101,Abbott_2017}, predict values of very roughly $a\approx 0.7$ \cite{PhysRevD.78.044002}. In these cases, the effect of the spin has to be taken into account and could potentially change the nonspinning result significantly. We address this issue in the following.
\par
Along with the area, we must now also quantize the spin, as usual,
\begin{align}
J = \hbar j,
\end{align}
where $j$ is a half-integer or integer. We can define the transition frequency $\omega$ for $n \rightarrow n - \delta n$ and $j\rightarrow j - \delta j$ to be
\begin{align}
\omega = \frac{c^2}{\hbar} \left(M(n, j) - M(n - \delta n, j - \delta j) \right).
\end{align}
The linearized version of this can be found in \cite{Foit:2016uxn}.
From the condition
\begin{align}
j = \frac{a G M(a,n)^2}{\hbar c},
\end{align}
one can then find $j(a)$ explicitly as a function of $n$ and $\alpha$. Next we use the results of Page \cite{Page:1976ki} (removing the neutrinos), which we can then match to the rates of change of area and angular momentum. Introducing 
\begin{align}
f&\equiv -M^2 \frac{\text{d}M}{\text{d}t} = -M^2 \dot{M},
\\
g&\equiv - \frac{M}{a} \frac{\text{d}J}{\text{d}t} = - \frac{M}{a} \dot{J},
\end{align}
one finds for the rate of area change

\begin{align}\label{Adot}
\dot{A} &= \frac{h c^4}{G^2} \frac{A}{M^3}  \left(\frac{g-2f}{\sqrt{1-a^2}} -g\right).\\
\end{align}
We can then use these expressions to find the semiclassical prediction (in this case for the rate of change of the area quantum number, i.e. $\dot{n}$) and match this to the quantum version in the same way as has been done for the nonspinning case like so,
\begin{align}
\dot{n} = -\frac{\left< \delta n \right> }{\tau},
\end{align}
which gives the width
\begin{align}
\Gamma = -\frac{\hbar \dot{n}}{\left<\delta n \right>} = -\frac{\hbar \dot{n}}{r \left<\delta n_0 \right>},
\end{align}
where we parametrized $\left<\delta n(j) \right>$ with $r$ to match the nonspinning case $j=0$ for $r=1$. 
We expect $r$ should be $\lesssim 1$, because for $a$ slightly larger than $0.8$, $r$ must go negative, due to the effect of the superradiance lead to the area growing (as mentioned in the Introduction). Furthermore, as there are more decay channels due to the spin, one could expect the lifetime of these states to be considerably lower than in the nonspinning case.
If we additionally define
\begin{align}
\omega_0 \equiv \omega (n\rightarrow n - 1, j=0),\\
\omega_\text{ref} \equiv \omega (n\rightarrow n - 1, j\rightarrow j),
\end{align}
we can finally compare the influence of the spin by looking at the ratio
\begin{align}\label{result}
\frac{\Gamma \omega_0}{\Gamma_0 \omega_\text{ref}}  = \frac{F(a)}{r},
\end{align}
again noting that $r$ and $F(a)$ must become negatice for large $a$. In the particularly interesting range of \(0.6<a<0.8\), \(F(a)\) ranges from \(\sim 3.5\) to \(\sim 5.5\). This means that, for the lowest value of \(\alpha\)(\(= 4 \ln 2\)), the \textit{natural} linewidth can cause neighboring states to completely overlap. It seems likely, once one includes the effects of the black hole not being in a specific stationary state, the incoming radiation being thermal, and other environmental effects, that only for considerably larger values of \(\alpha\) is it plausible that neighboring states are well separated. In our calculation we assumed that Hawking radiation gives the correct prediction for the rate of change of the black hole parameters. Note that a qualitatively similar conclusion for the widths was found in a recent work which instead relies on the rate of particle emission \cite{Hod:2015qfc}.
\par
\begin{figure}
\centering
\includegraphics[width=1.0\linewidth]{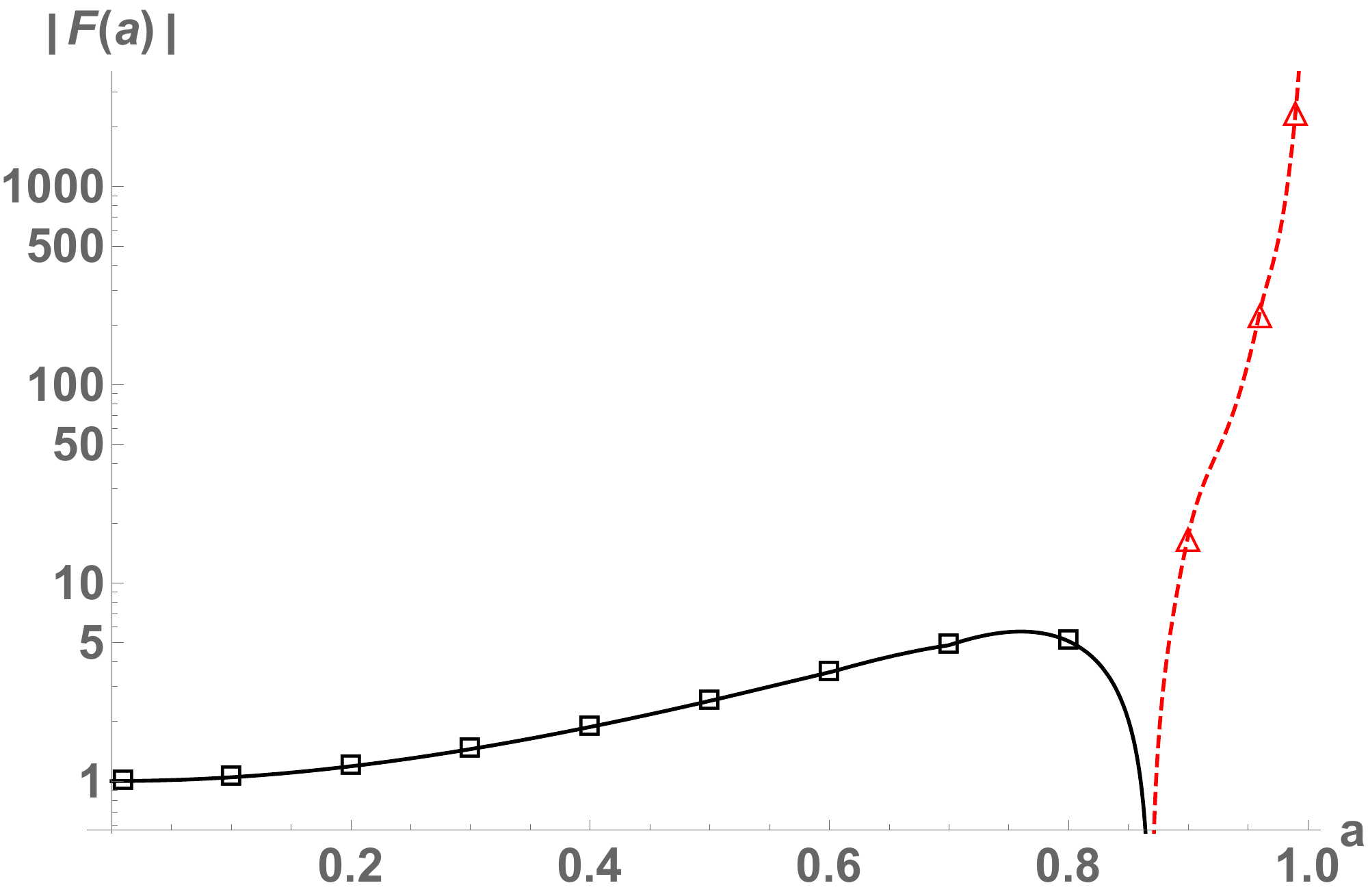}
\caption{Here we show the absolute value of the enhancement factor $F(a)$, see Eq. \eqref{result}, as a function of the spin. The black color represents positive values, whereas red is negative. The squares and triangles correspond to values of $a$ included in Table 1 in Ref. \cite{Page:1976ki}, while the lines are interpolations for demonstrative purposes. \label{figure_result}}
\end{figure}

Using  \(\dot{J}\) we can find the ratio of \(\left<\delta j\right>\) to \(\left<\delta n\right>\). For the same range, it varies from \(\sim 0.1 \alpha\) to \(\sim 0.4 \alpha\). If we expect the superradiant effects are not too strong (at least for the lower end of the interesting range) then one may still expect $\left<\delta j\right>\sim 1$, which is true for most of the range $4 \ln 2 < \alpha < 32\pi$ at $a=0.6$. We show $F(a)$ in Fig. \ref{figure_result}, where one finds the maximum around postmerger spins, while it changes sign for rapidly rotating black holes. That this must turn negative can be seen from Fig. 8 of Ref. \cite{Page:1976ki} and is a consequence of superradiance.
\section{Discussion}\label{Discussion}
Because of the lack of a fundamental theoretical framework, to estimate the width of the states one must make various assumptions: (0) area quantization is valid, (1) the energy of any given state is just its ADM mass, (2) the power output of spontaneous decays matches those of Hawking radiation, and (3) decays that significantly spin up the black hole are strongly suppressed. By including spin, we conclude that the states of black holes of astrophysical interest are very likely to strongly overlap. This is based on commonly proposed values for $\alpha$. If this value is much larger, the states of astrophysical black holes could be well separated, but how this can be justified is unclear.
\par
At this point, we want to connect our results to the expected reflectivity of astrophysical black holes. The correct method to do this for high intensity classical waves is unclear. In \cite{Oshita:2019sat} it is claimed that the reflectivity is independent of the spacing and widths of the states and described by a Boltzmann distribution. They arrive at this conclusion via three different calculations; however, the assumptions used there seem to require additional justification. The first way they obtain their result is using the classical Einstein rate equations of a pure two-level system with well-separated states in thermal equilibrium. A solar mass black hole is in an extremely high state ($n\sim 10^{76}$) and the nearby states are all evenly spaced. Additionally, we have seen that there is likely significant overlap between them. Their second approach uses a near horizon approximation, whether this avoids trans-Planckian frequencies is unclear to us, if it does not, including them would not be uncontroversial \cite{Jacobson:1999zk,Booth:2018xvb}. That the dissipation should appear so simply in the tortoise coordinate in particular also seems to require some more motivation. Their final calculation imposes $CP$ symmetry and leads directly to topology changes. Given that we know $CP$ symmetry is not realized in nature, it seems odd to apply this to an extremely nonlinear quantum gravity state. Recovering this infrared, approximate symmetry in the quantum gravity regime seems unlikely. Taking the result that occupations are well described by a Boltzmann distribution at face value, the most occupied state would be the ``minimal black hole'' state, which would be \(\sim e^{10^{78}}\) times more occupied than the solar mass black hole state.
\par
We want to make one last comment on the prediction of absorption lines \cite{Cardoso:2019apo}. The pure quantum mechanical treatment of black holes, which predicts a discrete spectrum, does not include the influence of the effective potential for perturbations. This potential, as shown in the same work, will trap perturbations up to the light ring. It is known that the trapped frequencies are low, while high frequency waves are only once partially being reflected when approaching the would-be horizon. However, the series of echoes at later times is built up from the frequencies of the trapped-quasinormal modes (QNMs) and is therefore discrete, see \cite{Macedo:2018yoi} for a recent analysis. In order to see absorption lines from area quantization in the echo signal, the QNM frequencies have to match the black hole state frequencies. This is, in general, not the case, which might make absorption hard to detect. However, the first reflection that includes the high frequency emission is more promising to look for absorption lines, because they are less influenced by the presence of the potential barrier.
\par
Finally, we have seen that astrophysical black holes are unlikely to reflect in the most common area quantization scenarios. Whether any quantum gravity candidates can support values of $\alpha$ that allow significant reflectivity remains open and is worth further investigation.
\acknowledgments
The authors thank the anonymous referees, as well as Vitor Cardoso and Qingwen Wang for valuable feedback. S. H. V. thanks Philip Wolf for useful discussions and was supported by the PhD scholarship Landesgraduiertenf\"orderung. The authors acknowledge support from the COST Action GW-verse CA16104.
\bibliography{literature}
\end{document}